\documentclass[%
 reprint,
 amsmath,amssymb,
 aps,
]{revtex4-2}
\usepackage{xcolor}
\usepackage{soul}
\usepackage[version=3]{mhchem}
\usepackage{graphicx}
\usepackage{dcolumn}
\usepackage{bm}

\begin{document}

\preprint{APS/123-QED}

\title{Impact of the Energy Landscape on the Ionic Transport of Disordered Rocksalt Cathodes}

\author{Shashwat Anand}
\author{Tina Chen}%
\author{Gerbrand Ceder}%
 \email{gceder@berkeley.edu}
\affiliation{Materials Sciences Division, Lawrence Berkeley National Laboratory, Berkeley,
CA 94720, USA}%

\date{\today}

\begin{abstract}
Traditional approaches to identify ion-transport pathways often presume equal probability of occupying all hopping sites and focus entirely on finding the lowest migration barrier channels between them. Although this strategy has been applied successfully to solid-state Li battery materials, which historically have mostly been ordered frameworks, in the emerging class of disordered electrode materials some Li-sites can be significantly more stable than others due to a varied distribution of transition metal (TM) environments. Using kinetic Monte Carlo simulations, we show that in such cation-disordered compounds only a fraction of the Li-sites connected by the so-called low-barrier ``0-TM" channels actually participate in Li-diffusion. The Li-diffusion behavior through these sites, which are determined primarily by the voltage applied during Li-extraction, can be captured using an effective migration barrier larger than that of the 0-TM barrier itself. The suppressed percolation due to cation disorder can decrease the ionic diffusion coefficient at room temperature by over 2 orders of magnitude.
\end{abstract}

\maketitle


Until recently, candidate materials studied for Li battery electrode applications have predominantly been ordered compounds.\cite{Meng_Van_der_Ven_2017_AEM_Review} A separation of sublattices on which Li and transition metal (TM) reside was considered essential for unhindered ionic transport in electrode materials. Commercialized layered cathodes --- an example of such ordered materials ---  therefore use only a few TMs (e.g. Ni, Co and Mn) which have the proper electronic structure to accommodate and retain the layered ordering for hundreds of cycles.\cite{Julia_2021_Molecules_Review} This constraint on the choice of metals has set the Li-ion battery industry on a path to consume a sizable fraction of the world's annual Co/ Ni production if energy storage goals of $>$ 1 TWh/year are to be met.\cite{Olivetti2017Joule_Co_ResourceProblem,USGeologicalSurvey2011mineral}

The recent development of Li-excess cation-disordered rocksalt (DRX) cathodes\cite{Lee_2014_Science_DRX, Yabuuchi2015_PNAS_LiNb,Chen2015_AEM_LiVF_DRX,AlexUrban2016_AEM_DRX,Kan2018_ChemMater_LMNO}, in which Li and other metals are disordered over the octahedrally coordinated sites in a rocksalt structure, lifts the restriction on a specific Li-TM ordering to enable the exploration of cathode materials in a much wider chemical space. In general, DRX cathodes are known to show high capacity\cite{Lee2018Mn_DoubleRedox_Nature,Lee2015_LNTMo_EES} and good stability.\cite{li2021fluorination_AFM_SurfaceStability,lee2017_Fluorine_HighCapa_NatComm,Raphael2020_DRX_EES-Review} Some DRX compounds\cite{Lun2021_NatMater_HE-DRX,Jianping_2021_NonTopotactic_NatEnergy} as well as similar disordered materials with partial spinel-like features\cite{Ahn_AEM_2022_SpinelLikeRearrangement} have also shown exceptional rate capability. DRX compounds are also being studied as potential anode materials.\cite{liu_DRXAnode_2020_Nature} But this evolution of electrode materials towards disordered materials necessitates the development of new models\cite{ML_NEB_2021_Vegge_Arghya} to understand the more complex ion transport through a disordered environment. While the basic transport theory in cation disordered rocksalts has been developed through percolation models, we show in this paper that a more rigorous understanding requires consideration of the varying site energy landscape of the Li-sites, and that in general this effect leads to a substantial reduction in Li-ion mobility.

\begin{figure*}[hbt!]
\centering
\includegraphics[scale=0.25]{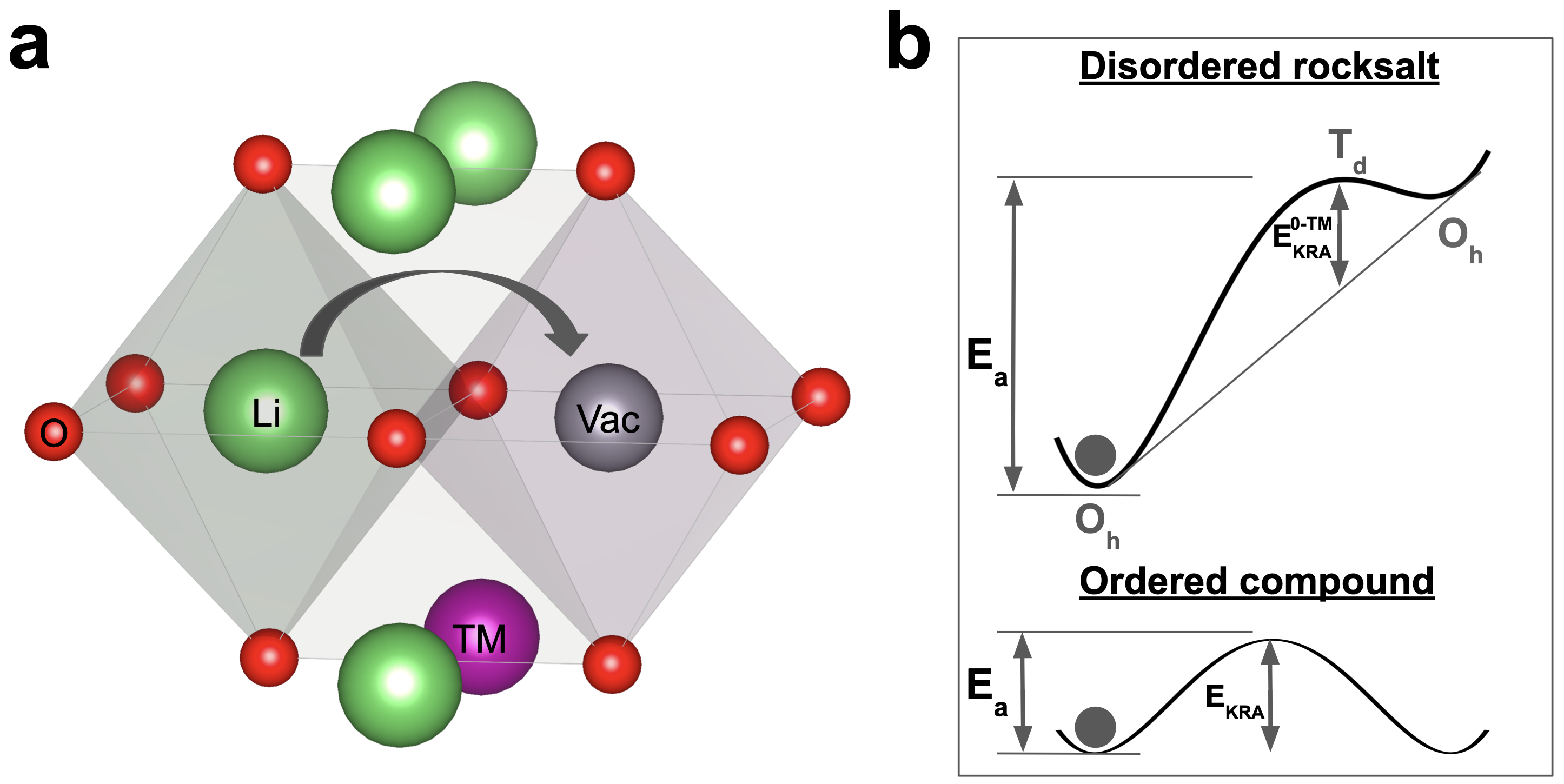}
\caption{(a) 3D view of 0-TM and 1-TM tetrahedral diffusion channels between adjacent edge sharing Li-sites in DRX compounds. The lower barrier 0-TM channel (black arrow) forms the basis of the percolating Li pathway. (b) Activation barrier E$_{\text{a}}$ associated with hopping out of an occupied octahedral
(O$_{\text{h}}$) site in a DRX system versus a typical ordered compound. Even with the same kinetically resolved activation barrier (E$_{\text{KRA}}$), E$_{\text{a}}$ in DRX systems can be much larger due to a difference in the O$_{\text{h}}$ site energies.}
\label{fig:Landscape_VoltageProfile}
\end{figure*} 

In close-packed oxides, the Li-ion resides in the oxygen octahedra (O$_\text{h}$) and migrates through intermediate face-sharing tetrahedral sites (T$_\text{d}$). In DRX cathodes, Li migration occurs through low-barrier Li-rich tetrahedral environments (0-TM channels, where the tetrahedral site does not face-share with any transition metal or TM) as this provides a low electrostatic repulsion pathway for migration (Figure \ref{fig:Landscape_VoltageProfile} a). Therefore, Li-excess compositions ($x > 0.1$ in \ce{Li_{1+$x$}TM_{1-$x$}O_2}) in which all the 0-TM channels are connected as a percolating network\cite{Urban_2014_RX_Config_AEM} are expected to boost rate capability and maximize reversible capacity. Compared to 1-TM channels for diffusion, through which Li-migration in layered cathodes occurs, the migration barriers for 0-TM channels in DRX compounds are generally much smaller and should lead to ionic diffusion at least two orders of magnitude faster.\cite{Lee_2014_Science_DRX} Despite this, measured values of Li-diffusivity in DRX compounds (10$^{-15}$ cm$^2$ s$^{-1}$)\cite{Huiwen2019_NatComm_LMT(Z)O,Jianping_2021_NonTopotactic_NatEnergy,lee2021_DRX_diffusion} are often a few orders of magnitude smaller than that of layered compounds (10$^{-8}$ cm$^{2}$ s$^{-1}$ - 10$^{-13}$ cm$^2$ s$^{-1}$) with occasional measurements\cite{chen2015LVO_DiffusionData} reported in a comparable range. Barring a few exceptions, Li-excess DRX cathodes mostly show limited rate capability, which may be connected to poor Li-ion diffusion. Therefore, the current understanding of Li transport based entirely on 0-TM percolation is used largely only for qualitative comparisons of Li diffusion and rate capability between compounds. \cite{Lun2021_NatMater_HE-DRX, Huiwen2019_NatComm_LMT(Z)O} 


The traditional percolation theory focuses on understanding Li-transport pathways by ensuring connectivity of the low energy intermediate T$_\text{d}$ sites. However, the cation disorder in DRX compounds also lead to a varied TM environment, which influences the energies of Li O$_\text{h}$ sites. Li transport therefore will not depend only on the (T$_\text{d}$) site environments (0-TM, 1-TM and 2-TM) but also on local variations in site energy. The sketch in Figure \ref{fig:Landscape_VoltageProfile} b contrasts the migration barrier ($E_{\text{a}}$) associated with a Li hop in the DRX compounds with that in ordered compounds in the dilute carrier limit. While in ordered compounds the surrounding TM environments of the Li sites are mostly the same, the site energies of the O$_\text{h}$ Li sites in the DRX compounds vary causing the migration barrier to change depending on the initial and final positions of the Li-ion. As a result of such a varied energy landscape, the migration barrier of 0-TM channels in DRX compounds can vary significantly and a percolation analysis based on 0-TM connectivity alone --- although applicable to ordered counterparts --- may not be sufficient. For larger variations in site energy, some sites cannot be occupied, limiting the Li-sites in the 0-TM percolating network that can participate in ionic transport.

The effect of varying site energies on Li-transport can be isolated by defining a kinetically resolved activation (KRA) barrier $E_{\text{KRA}}$,\cite{VanderVen_2000_LiCoO2} which represents the contribution from the T$_\text{d}$ site environment ($E^{\text{0-TM}}_{\text{KRA}}$ for 0-TM channels) to the overall magnitude of the migration barrier $E_{\text{a}}$. $E^{\text{0-TM}}_{\text{KRA}}$ is obtained by subtracting the average of the energies for the end points of the hop from the energy at the activated state T$_\text{d}$ as shown in Figure \ref{fig:Landscape_VoltageProfile} b. Here, we study Li-transport in a model DRX system keeping $E^{\text{0-TM}}_{\text{KRA}}$ constant and varying the extent of variance in the Li site energies.

In this paper we perform kinetic Monte Carlo (KMC) simulations to calculate Li-diffusion within a simplified model which includes varying site energies, mimicking the Li-TM site disorder as well as nearest neighbor Li-Li interactions. We show that in disordered cathodes where the standard deviation in Li site energy distribution ($w_\text{SE}$) is much larger than the thermal energy ($k_{\text{B}}T$), traditional percolation theory is not sufficient for understanding Li transport. Due to the larger migration barriers associated with the escape from lower-energy sites, only a fraction of the Li-sites within the 0-TM percolating network actually participate in ionic transport at a given voltage (which sets the Li content and average energy of Li-ions). In real DRX materials we find that the magnitude of $w_\text{SE}$ can be comparable to $E^{\text{0-TM}}_{\text{KRA}}$ itself, and the impact of cation disorder in limiting Li percolation is sufficient to suppress the room temperature diffusion coefficient by over 2 orders of magnitude.

To investigate the effect of a disordered energy landscape on cation diffusion, we construct a model system with a single FCC lattice on which Li and vacancies (Vac) are the only two species allowed. The transport properties in this model system are calculated with varying Li occupancies ($x_{\text{Li}}$) in the range [0,1]. To mimick the interactions between Li and TM atoms in a system, the Li site energy is varied according to a Gaussian distribution with standard deviation $w_\text{SE}$. The Li site energy landscape was kept static during Li diffusion to simulate a fixed TM environment (topotactic process). The interaction between Li sites is limited in the model to nearest neighbors and is accounted for using the effective cluster interaction $J_\text{Li-Vac}^\text{NN}$ (see supplementary material)\cite{Asta_1991_PRB_CI-ECI}. Kinetic Monte Carlo simulations\cite{Andersen2019_KMC,VanderVen2001_LiCoO2_PRB} were performed using a rejection-free algorithm\cite{Gillespie1976_Rejection-free1,Gillespie1977_Rejection-free2} to determine the tracer diffusion coefficient ($D$) and the correlation factor $f$.

\begin{figure*}[hbt!]
\centering
\includegraphics[scale=0.25]{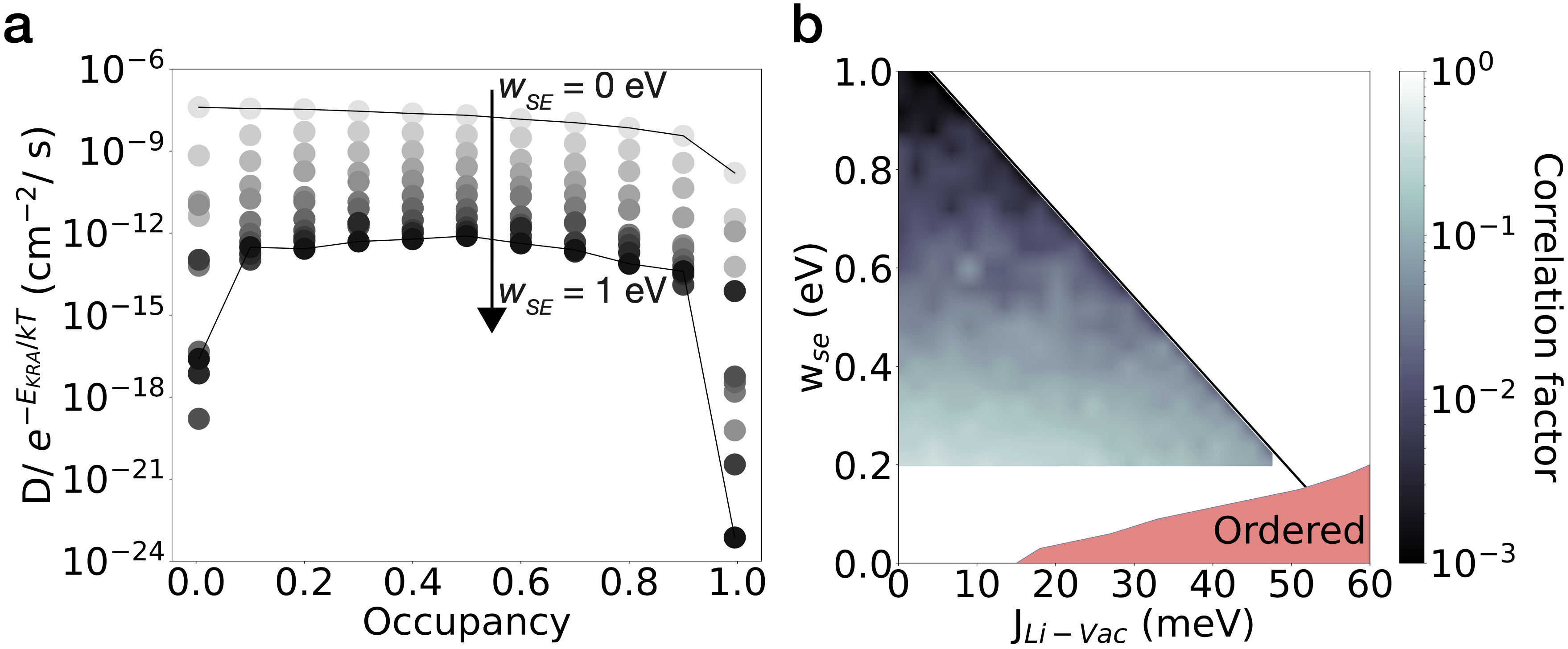}
\caption{(a) Impact of varying O$_\text{h}$ site-energy distribution (standard deviation $w_\text{SE}$) on the calculated Li diffusion coefficient at different occupancy of the O$_\text{h}$ sites. The magnitude of $w_\text{SE}$, represented by the shade of the data points, is varied over 11 values in the range [0 eV, 1 eV]. (b) Contour plot of calculated correlation factor (f) at $x_\text{Li}$ = 0.3 in the 2-dimensional parameter space ($w_\text{SE}$, $J_\text{Li-Vac}^\text{NN}$) of the model system. The parameters which best fit the experimentally measured voltage profile of \ce{Li_{1.17-x}Mn_{0.343}V_{0.486}O_{1.8}F_{0.2}} are given by the solid black line. The region in which the model system becomes ordered at room temperature (T = 300 K) is shaded in salmon. }
\label{fig:Diffusion_CorrelationFactor}
\end{figure*}

To isolate the effect of the varying O$_\text{h}$ site energy distribution on Li-diffusion, the tracer diffusion coefficient was calculated for $J_\text{Li-Vac}^\text{NN} = 0$ and constant $E^{\text{0-TM}}_{\text{KRA}}$ with $w_\text{SE}$ varied in the range [0 eV, 1 eV]. The calculated $D$ at different Li occupancies is shown in Figure \ref{fig:Diffusion_CorrelationFactor}a. Data corresponding to larger $w_\text{SE}$ values are shaded with darker symbols and data for the extreme $w_\text{SE}$ values are connected by solid lines. In general, we find that the variance in O$_\text{h}$ site energy distribution suppresses the diffusion coefficient at all Li compositions. In the non-dilute regions, the diffusion coefficient decreases by over 4 orders of magnitude for $w_\text{SE}$ = 1 eV in comparison to the ordered compound ($w_\text{SE}$ = 0 eV). The impact of cation disorder is much more pronounced in the dilute limit where $D$ is reduced by over 10 orders of magnitude at $w_\text{SE} \sim 1$ eV. 

To gain insights into the origin of the large drop in $D$ with $w_\text{SE}$, we inspect the corresponding calculated correlation factor ($f$), which is a measure of the degree to which successive hops of the diffusing species are correlated ($f$ = 1 for non-interacting particles performing random walk).\cite{VanderVen_2020_BatteryComputationalReview_ChemReviews,Kolli-VanderVen_2021_ChemMater_MgTiS2} In variants of the rock-salt compounds with layered cation ordering such as Li$_x$CoO$_2$, \textit{f} is typically $>$ 10$^{-1}$, except at values of $x$ approaching 1.\cite{VanderVen2001_LiCoO2_PRB} Calculated $f \sim$ 1 for $w_\text{SE}$ = 0 eV and $J_\text{Li-Vac}^\text{NN}$ = 0 meV in the model signify an unhindered random walk because successive Li-hops have no correlation between them. In Figure \ref{fig:Diffusion_CorrelationFactor} b, the correlation factor as a function of $w_\text{SE}$ and $J_\text{Li-Vac}^\text{NN}$ at $x_{\text{Li}}$ = 0.3 and $T$ = 300 K using a color scale. The region corresponding to ordered Li-Vac configurations in the model is shaded salmon (see supplementary information) and is not relevant here. The solid line exemplifies the possible relative magnitudes of $w_\text{SE}$ and $J_\text{Li-Vac}^\text{NN}$ in real DRX systems as estimated by comparing the calculated voltage profile for the model with the experimentally measured voltage profile for \ce{Li_{1.17-x}Mn_{0.343}V_{0.486}O_{1.8}F_{0.2}} (see discussion below). We find that by increasing $w_\text{SE}$ alone, as in Figure \ref{fig:Diffusion_CorrelationFactor} a, \textit{f} is suppressed by three orders of magnitude, suggesting a strong correlation between successive hops. Although \textit{f} is suppressed with increasing $J_\text{Li-Vac}^\text{NN}$ as well, its effect on \textit{f} appears much weaker than that of $w_\text{SE}$. For $w_\text{SE} \sim 0.2$ eV, $f$ is of the same order of magnitude as Li$_x$CoO$_2$ at similar Li-concentration ($x$ = 0.3), suggesting that this extent of site energy variation does not have a very adverse impact on Li-diffusion. 

\begin{figure}[hb!]
\centering
\includegraphics[scale=0.15]{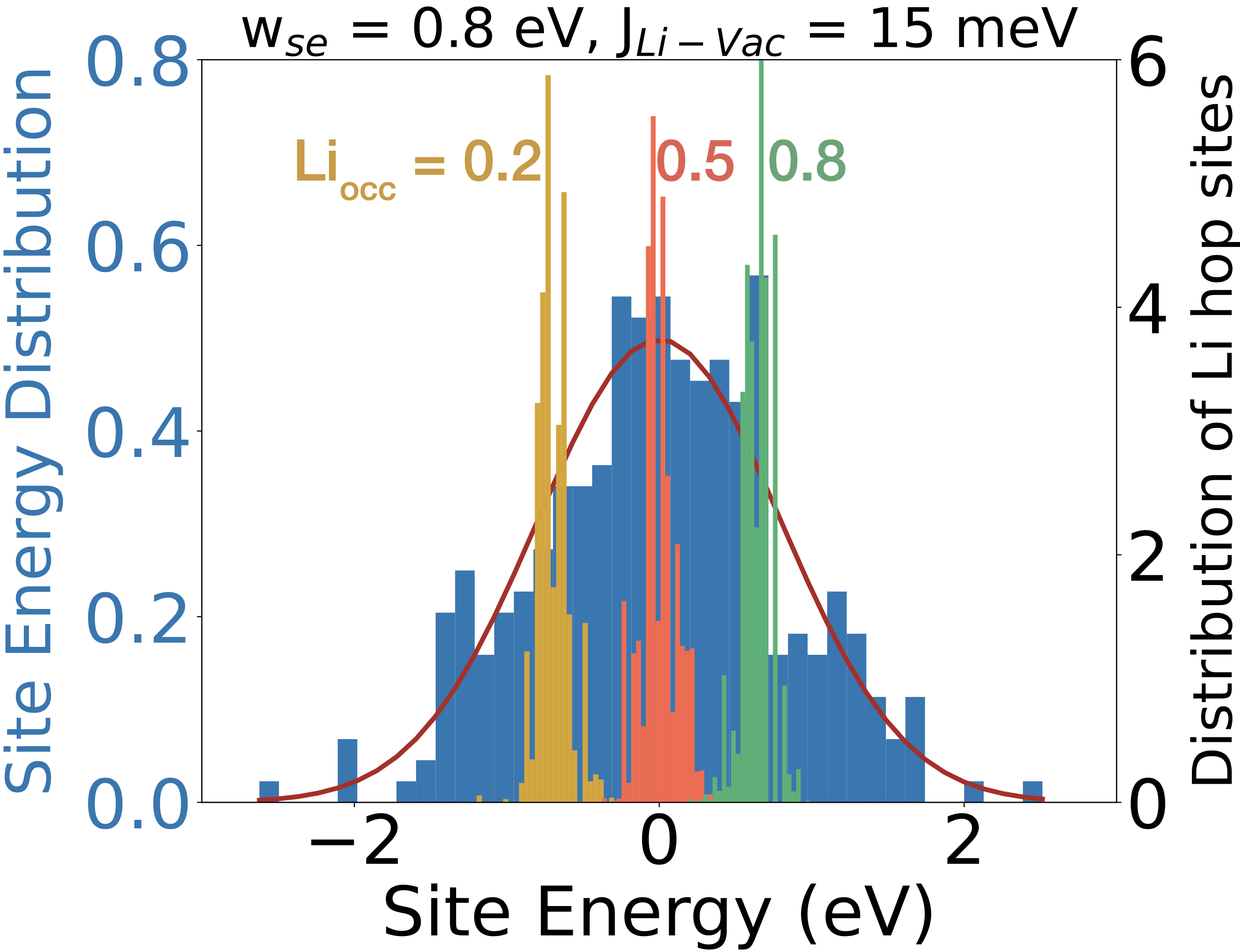}
\caption{Li site energy distribution (blue histogram) with standard deviation $w_\text{SE}$ = 0.8 eV. Distribution of Li sites involved in Li-diffusion at $x_\text{Li}$ = 0.2, 0.5, 0.8.}
\label{fig:Percolation_Summary}
\end{figure}

To understand the impact of cation disorder on Li site percolation, we compare in Figure \ref{fig:Percolation_Summary} the site energy distribution of all Li sites (blue histogram, left y-axis) against the energy distribution of those sites that actually participate in diffusion (yellow, orange, green histograms, right y-axis) at room temperature for different values of $x_\text{Li}$. The solid red curve represents the Gaussian distribution curve with the same standard deviation as the $w_\text{SE}$ (= 0.8 eV) used in the model. We find that most of the transport occurs predominantly through Li sites with site energies falling within a few k$_\text{B}T$ of the highest energy occupied Li site. All vacant sites above this range and occupied sites below it are effectively immobile due to the prohibitively large migration barriers connecting nearest-neighbors with relatively larger site energy differences. For the model system with $w_\text{SE}$ = 0.8 eV, the sites participating in transport for $x_\text{Li}$ = 0.2, 0.5 and 0.8 are almost entirely different from each other. These results show that during delithiation, the sites available for transport will depend on the voltage applied and can be severely limited when $w_\text{SE} >> k_\text{B}T$. As a result, the fraction of Li sites available for percolation can effectively be much smaller than that expected from the 0-TM channel connectivity\cite{Urban_2014_RX_Config_AEM} for a given $x$ in Li$_x$TM$_{1-x}$O$_2$. 

\begin{figure}[hbt!]
\centering
\includegraphics[scale=0.2]{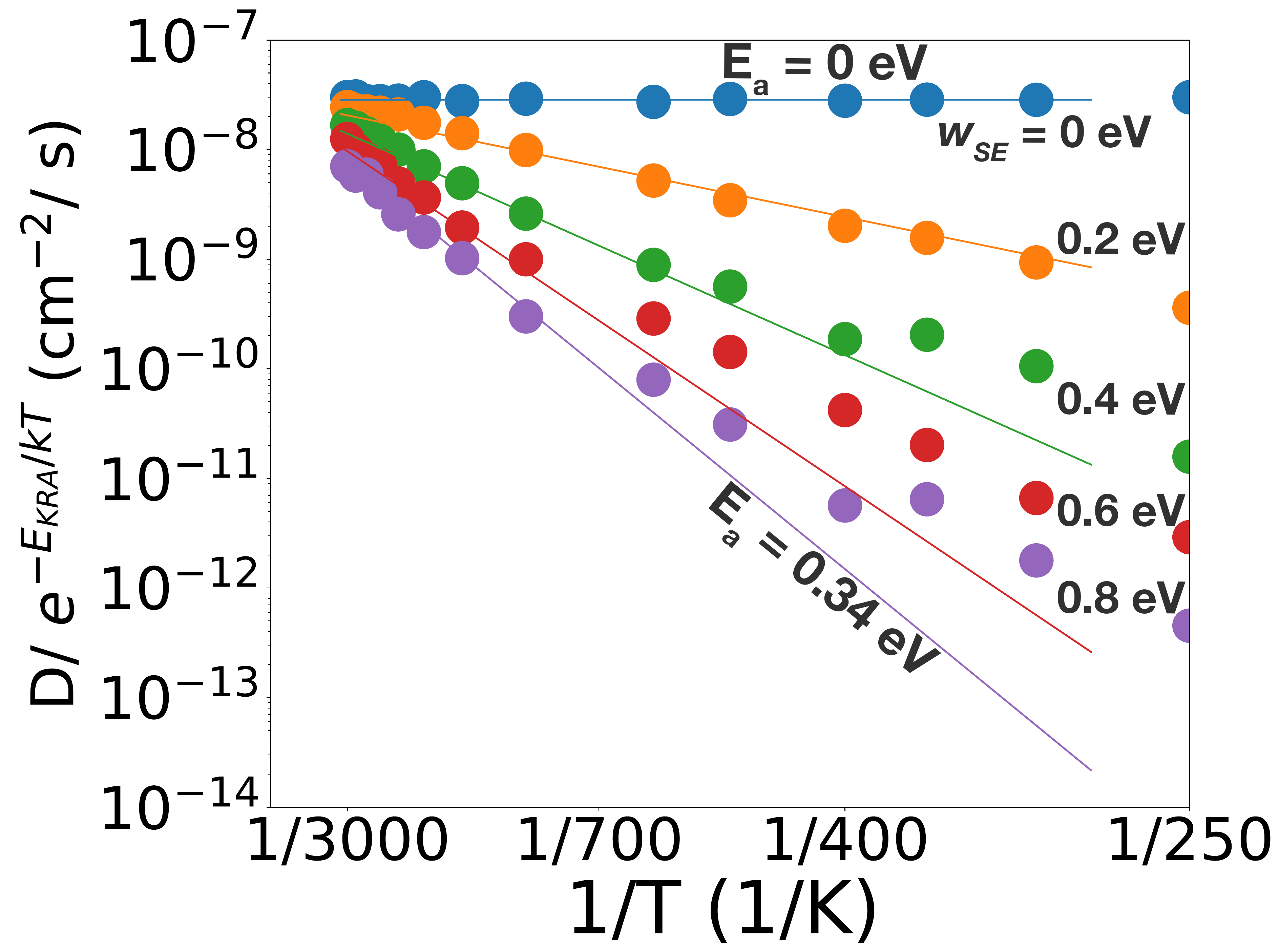}
\caption{Arrhenius plot of calculated tracer diffusion coefficient for model DRX systems at $x_\text{Li}$ = 0.3 with varying $w_\text{SE}$ in the range [0, 0.8]. Corresponding high temperature data is fitted by an increasing effective activation barriers of magnitude $\sim E_\text{KRA}$+0.45 $w_\text{SE}$.}
\label{fig:EffectiveActivationBarrier}
\end{figure}

To capture the effect of the varied O$_\text{h}$ energy landscape on the overall Li diffusion, an Arrhenius plot of $D$ for multiple $w_\text{SE}$ is shown in Figure \ref{fig:EffectiveActivationBarrier} (for $x_{\text{Li}}$ = 0.3). Note that values of $D$ are normalized by $exp\left(\frac{-\text{E}_\text{KRA}}{\text{k}_\text{B} T} \right)$ to isolate the effect of the site energy distribution on the activation energy. Consistent with the calculations for room temperature in Figure \ref{fig:Diffusion_CorrelationFactor}, $D$ is suppressed at all temperatures with increasing $w_\text{SE}$. In the high temperature limit, $D$ displays an Arrhenius behavior varying as $D \propto exp\left(\frac{-\text{E}_\text{a}}{\text{k}_\text{B} T} \right)$ and suggests that the transport through 0-TM channels can still be modeled with a single \textit{effective} activation barrier ($E_{\text{a}}^{\text{eff}}$), which approximately behaves as $E_{\text{a}}^{\text{eff}} \sim E_{\text{KRA}} + 0.45 w_\text{SE}$. At lower temperatures, the diffusion coefficient gradually deviates from this behavior, signifying a smaller effective activation barrier caused by the participation of sites lying in a narrower energy window. Importantly, Figure \ref{fig:EffectiveActivationBarrier} shows that, unlike in ordered compounds, $E_{\text{a}}$ $\ne$ $E_{\text{KRA}}$ in disordered materials even when nearest neighbor Li-Vac interactions are absent ($J_\text{Li-Vac}^\text{NN}$ = 0 eV). Furthermore, results in Figure \ref{fig:EffectiveActivationBarrier} suggest that $E_{\text{a}}^{\text{eff}} > E_{\text{KRA}}^{0-TM}$ and $E_{\text{a}}^{\text{eff}}$ lies quite close in range to $E_{\text{KRA}}^{1-TM}$ (typically $E_{\text{KRA}}^{1-TM}$ - $E_{\text{KRA}}^{0-TM} \sim$ 0.2 eV \cite{Raphael2020_DRX_EES-Review}), and 1-TM channels could therefore also participate in ionic-transport.  


\begin{figure}[hbt!]
\centering
\includegraphics[scale=0.25]{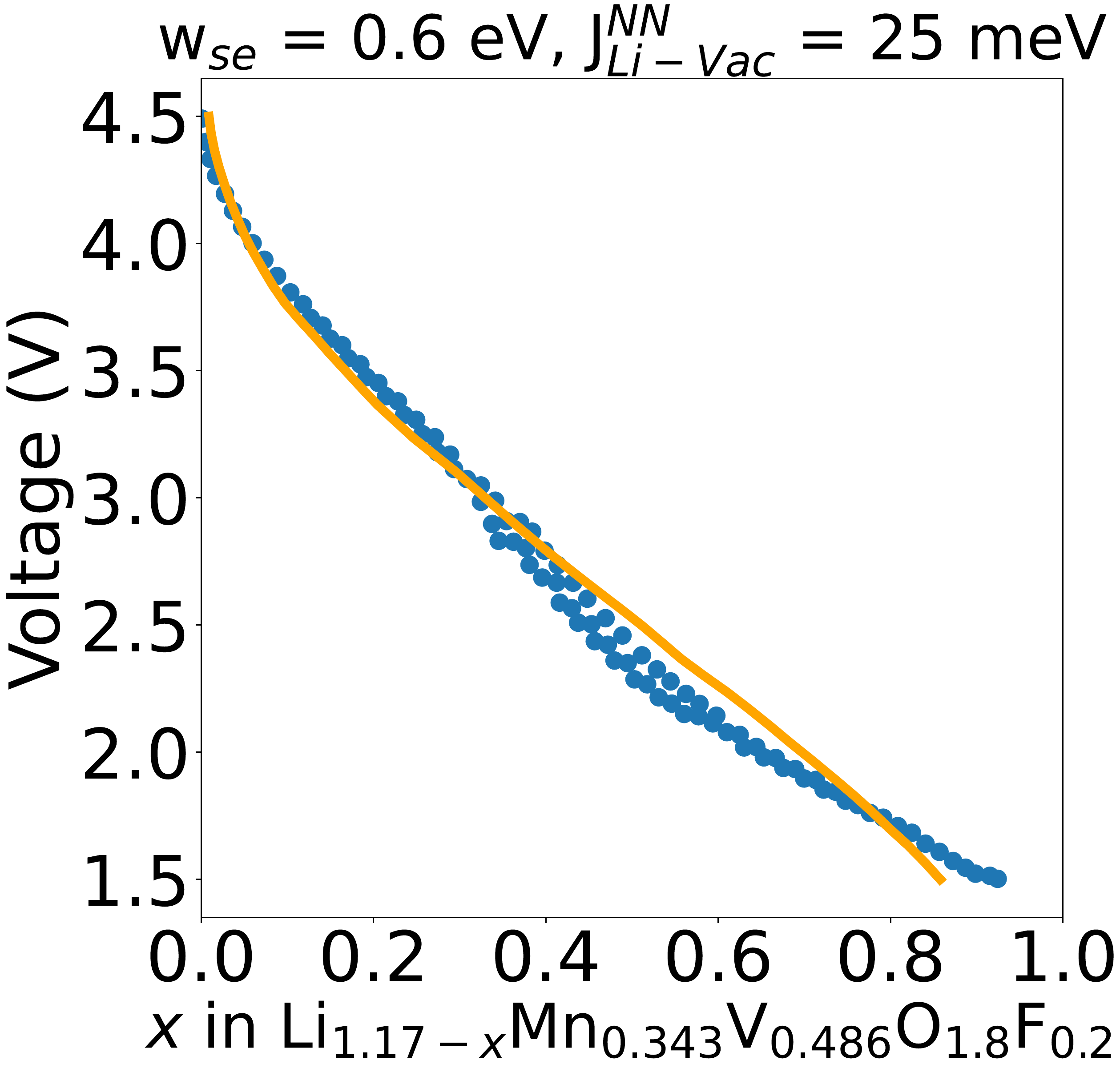}
\caption{Calculated voltage profile for model DRX system with $w_\text{SE}$ = 0.6 eV and $J_\text{Li-Vac}^\text{NN}$ = 25 meV fit to the experimentally measured voltage profile of \ce{Li_{1.17-x}Mn_{0.343}V_{0.486}O_{1.8}F_{0.2}}. }
\label{fig:VoltageProfile_Comparision}
\end{figure}

The typical value for $J_\text{Li-Vac}^\text{NN}$ used in our model can be expected to be similar to that of the layered variants of the rock-salt type crystal structure, where $J_\text{Li-Vac}^\text{NN}$ $\sim$ 15-35 meV \cite{VanderVen1998_LiCoO2_PRB,Dompablo_2002_LiNiO2_PRB}. Physically reasonable values of $w_\text{SE}$ in DRX compounds can then be gauged by comparing the model's calculated voltage profile to experimentally measured voltage profiles. Figure \ref{fig:VoltageProfile_Comparision} shows an example for such a comparison of the \ce{Li_{1.17-x}Mn_{0.343}V_{0.486}O_{1.8}F_{0.2}} experimental profile with the calculated voltage profile at $w_\text{SE}$ = 0.6 eV and $J_\text{Li-Vac}^\text{NN}$ = 25 meV. Because both $J_\text{Li-Vac}^\text{NN}$ and $w_\text{SE}$ contribute to the slope of the calculated voltage profile ($\frac{dV}{dx_{Li}}$),\cite{Abdellahi_2016_ChemMater_V_Profile} a series of values for them exist that will lead to the same voltage slope, which for \ce{Li_{1.17-x}Mn_{0.343}V_{0.486}O_{1.8}F_{0.2}} is shown in Figure \ref{fig:Diffusion_CorrelationFactor}b by the solid line. Since the slope of the \ce{Li_{1.17-x}Mn_{0.343}V_{0.486}O_{1.8}F_{0.2}} voltage profile is one of the largest among experimentally measured DRX compounds, the corresponding values of $w_\text{SE}$ and $J_\text{Li-Vac}^\text{NN}$ can be expected to be almost an upper limit for DRX compounds. We note that the Li-site energy landscape could also depend on variations in charge of the cation environment with Li-extraction (e.g. Mn$^{3+}$ $\rightarrow$ Mn$^{4+}$) as well as the redox chemistry of the transition metal involved. As a result, comparisons like the one shown in Figure \ref{fig:VoltageProfile_Comparision} can only provide approximate estimates of $w_\text{SE}$, and voltage profiles calculated with system-specific cluster expansions could provide more insight into the Li-site energy landscape of DRX compounds.\cite{VanderVen_2020_BatteryComputationalReview_ChemReviews,Tina_2018_MgSpinel_ChemMater}

Despite its approximate nature, our analysis can be used to develop general guidelines for correlating the rate capability of disordered cathode materials to their voltage profile. Since increasing either $J_\text{Li-Vac}^\text{NN}$ or $w_\text{SE}$ suppresses Li-diffusion and increases the slope of the voltage profile, one can expect rate capability to be better for materials with a flatter voltage profile. An example of this behavior can be found in partially disordered Li-excess spinel materials Li$_{1+x+y}$Mn$_{2-y}$O$_{4}$ for which the rate capability and voltage profiles were studied with varying $x$ and fixed Mn content.\cite{Zijian_2021_Matter_Order-to_Disorder_SpinelType} The rate capability was found to be better for compounds with relatively flatter voltage profiles, suggesting that the improved performance --- besides a larger fraction of kinetically accessible Li through 0-TM percolation --- could be caused by the smaller degree of disordering in them.

In addition to cation disorder, disorder on the anionic sub-lattice (F$^{-}$, O$^{-}$ and O$^{2-}$) due to the occurrence of O redox mechanism\cite{luo2016_NatChem_OxygenRedox,Hong2019_NatMater_OxygenRedox,Yabuuchi2016_NatComm_OxygenRedox,Freire_2016_NatMater_OxygenRedox,Freire_2017_JMCA_OxygenRedox} or fluorination\cite{Raphael2020_DRX_EES-Review,GChen_2020_AEM_Fluorination} may also effect ionic transport by modifying the the Li-site energy landscape. For instance, fluorination has been shown to increase the asymmetry of migration barriers in o-LiMnO$_2$ by over 100 meV due to changes in local Li site energy, although it had minimal effect on the $E_{\text{KRA}}$.\cite{Zinab_2020_JMCA_Fluorination_Migration_Barrier} Besides anion disorder, our analysis for DRX compounds could in principle also be extended to describe ionic transport in (i) compounds containing mobile TM elements such as Cr which undergo reversible octahedral-to-tetrahedral (oct–tet) migration during electrochemical cycling\cite{Jianping_2021_NonTopotactic_NatEnergy} and (ii) partially disordered spinel compounds\cite{Huiwen2020_PDS_NatEnergy}, where the T$_\text{d}$ and O$_\text{h}$ site energy distributions overlap to cause a rearrangement of Li-occupancy.

In conclusion, we specifically identify the effect of the varied TM environment seen by Li-sites in DRX compounds on ionic transport. This effect of Li-TM interactions is isolated by calculating diffusion coefficients for increasing variances in Li-site energy distributions and contrasted with that of Li-Vac interaction. We find that, for a DRX compound with a given voltage profile, the impact of Li-TM interactions on suppressing the diffusion coefficient can be much stronger than that of the Li-Vac interactions. Suppressed Li-diffusion occurs despite the connectivity of Li-sites through a network of 0-TM channels and results in part from effectively ``immobile'' Li-ions, which reside in sites with significantly lower energy than the Li-ions that participate in transport. Therefore, in comparison to simple ordered compounds, identifying ionic transport pathways in disordered compounds requires considering two additional levels of complexity namely: (a) connectivity through lowest barrier channels between hopping sites and (b) the varied energy landscape of the hopping sites itself.  

We would like to thank Zinab Jadidi for discussions on the subject of transport and Li-TM interactions in DRX compounds. This work was supported by the U.S. Department of Energy, Office of Science, Basic Energy Sciences, Materials Sciences and Engineering Division under Contract No. DE-AC02-05-CH11231 within the (GENESIS EFRC) program (DE-SC0019212). T.C. acknowledges financial support from the Assistant Secretary for Energy Efficiency and Renewable Energy, Vehicle Technologies Office, under the Applied Battery Materials Program, of the U.S. Department of Energy (DOE) under Contract No. DE-AC02-05CH11231. The computational analysis was performed using computational resources of the National Energy Research Scientific Computing Center (NERSC), a DOE Office of Science User Facility supported by the Office of Science of the US Department of Energy under contract no. DE-C02-05CH11231.


\bibliography{apssamp}


\end{document}